\documentclass[aps,prl,twocolumn,showpacs,groupedaddress]{revtex4}
\usepackage{graphicx}

\begin{document}


\title{Production of cold formaldehyde molecules for study and control of chemical reaction dynamics with hydroxyl radicals}



\author{Eric~R.~Hudson}
\affiliation{JILA, National Institute of Standards and Technology
and University of Colorado, Department of Physics, University of
Colorado, Boulder, CO 80309-0440 } \email[Electronic Address:
]{Eric.Hudson@Colorado.edu}

\author{Christopher~Ticknor}
\affiliation{JILA, National Institute of Standards and Technology
and University of Colorado, Department of Physics, University of
Colorado, Boulder, CO 80309-0440 }

\author{Brian~C.~Sawyer}
\affiliation{JILA, National Institute of Standards and Technology
and University of Colorado, Department of Physics, University of
Colorado, Boulder, CO 80309-0440 }

\author{Craig~A.~Taatjes}

\altaffiliation[JILA visiting fellow 2004. Permanent Address:
]{Combustion Research Facility, Sandia National Laboratories,
Livermore, CA 94551-0969 }

\author{H.~J.~Lewandowski}
\affiliation{JILA, National Institute of Standards and Technology
and University of Colorado, Department of Physics, University of
Colorado, Boulder, CO 80309-0440 }

\author{J.~R.~Bochinski}
\altaffiliation[Present Address: ]{Department of Physics, North
Carolina State University, Raleigh, NC 27695}

\author{John~L.~Bohn}
\affiliation{JILA, National Institute of Standards and Technology
and University of Colorado, Department of Physics, University of
Colorado, Boulder, CO 80309-0440 }

\author{Jun~Ye}
\affiliation{JILA, National Institute of Standards and Technology
and University of Colorado, Department of Physics, University of
Colorado, Boulder, CO 80309-0440 }


\date{\today}

\begin{abstract}
We propose a method for controlling a class of low temperature
chemical reactions.  Specifically, we show the hydrogen abstraction
channel in the reaction of formaldehyde (H$_{2}$CO) and the hydroxyl
radical (OH) can be controlled through either the molecular state or
an external electric field. We also outline experiments for
investigating and demonstrating control over this important
reaction.  To this end, we report the first Stark deceleration of
the H$_{2}$CO molecule. We have decelerated a molecular beam of
H$_{2}$CO essentially to rest, producing cold molecule packets at a
temperature of 100 mK with a few million molecules in the packet at
a density of $\sim$10$^{6}$ cm$^{-3}$.
\end{abstract}


\maketitle

Recent exciting developments in ultracold matter research include
the creation of ultracold molecules by magneto-association
\cite{Donley:2002}, leading to molecular Bose-Einstein condensation
\cite{BECBCS}. Despite the rich physics demonstrated in these
systems, all are characterized by spherically symmetric interactions
whose effect is included through one parameter, namely the s-wave
scattering length. By contrast, the permanent electric dipole moment
possessed by polar molecules permits long-range and anisotropic
interactions and enables new methods for external control in an
ultra-cold environment \cite{MagneticDipole}. The electric
dipole-dipole interaction (and control over it) gives rise to unique
physics and chemistry including novel collision and chemical
reaction dynamics. Lack of spherical symmetry in the interaction
causes colliding molecules to be attracted or repelled depending on
their relative orientation. Thus, an external electric field, which
orients the molecules, will have a profound effect on the molecular
interactions \cite{BOHN}. Furthermore, because the transition states
of a chemical reaction often involve specific orientations of the
molecular dipoles, an external electric field may shift the energy
barrier to reaction, making a particular reaction pathway more or
less favorable. A Stark decelerator
\cite{Bethlem:1999,Bochinski:2003,Bochinski:2004} producing cold
polar molecules with tunable and well-defined translational energy
is thus an ideal tool for the study of low (or negative) barrier
chemical reactions. We note other experiments utilizing
photo-association \cite{Wang:2003,Kerman:RbCs2004,Mancini:2004} have
succeeded in producing ultra-cold heteronuclear alkali dimers.

In this letter, we present a study of the collision and reaction
properties of OH with H$_{2}$CO at low temperatures, followed by a
detailed description of experiments required and carried out so far
to probe these novel dynamics. The calculations suggest for the
first time that chemical reactions, as well as collision cross
sections, can be altered orders of magnitude by varying either the
molecular state, or external electric field strength. We
specifically consider the H-abstraction channel in the reaction of
H$_{2}$CO and OH: $H_{2}CO + OH \rightarrow CHO + H_2O$. This
reaction not only represents a key component in the combustion of
hydrocarbons, but also plays an important role in atmospheric
chemistry where it is the primary process responsible for the
removal of the pollutant, H$_{2}$CO
\cite{Morris:1971,Soto:1990,Stief:1980,Dupuis:1984,Niki:1984,Yetter:1989,Li:2004}.
Near room temperature the rate of this reaction is weakly dependent
on temperature, suggesting the barrier to the process (if any) is
very low \cite{Stief:1980}. Measurements of the thermal activation
energy, $E_a$, are inconclusive
\cite{Morris:1971,Soto:1990,Stief:1980,Dupuis:1984,Niki:1984,Yetter:1989,Sivakumaran:2003},
ranging from $E_a/R =$ 750 K to -931 K, with the most recent
measurement giving a value of -135 K \cite{Sivakumaran:2003}. Here
$R$ is the universal gas constant. The most accurate calculations
\cite{Danna:2003} predict the energy of the transition state for the
abstraction to lie between -700 K and +60 K relative to the
reactants.

Experimentally, both OH
\cite{Bochinski:2003,Bochinski:2004,Bas:OH:2005} and H$_{2}$CO (this
work) molecules have been produced at low temperatures via Stark
deceleration. In the case of spontaneous reaction ($E_a<0$)
\cite{Morris:1971,Stief:1980,Niki:1984,Yetter:1989}, by magnetically
trapping OH in the presence of a tunable bias electric field and
``bombarding'' the trap with decelerated H$_{2}$CO packets the OH -
H$_{2}$CO total scattering and reaction rates can be mapped as a
function of collision energy and applied electric field. Thus $E_a$
can be measured. If $E_a > 0$, as some measurements and theory
suggest \cite{Soto:1990,Dupuis:1984,Li:2004,Danna:2003}, the
collision energy tuning afforded by the Stark decelerator provides a
direct way to measure the energy barrier to reaction. Unlike thermal
kinetics studies, which rely on fitting the Arrhenius formula to
reaction rates, the Stark decelerator can be used to tune the
collision energy above and below the threshold energy.

\begin{figure}
\resizebox{1\columnwidth}{!}{%
  \includegraphics{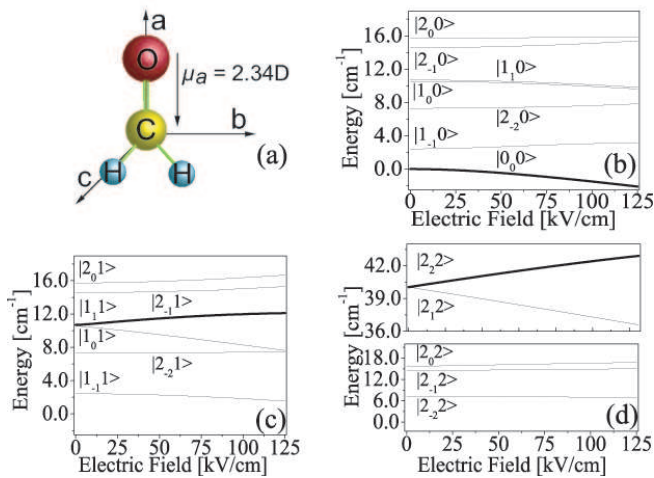}
} \caption{(a) (Color online) H$_{2}$CO. (b-c) Stark shifts of the
low-lying rotational states of H$_{2}$CO. States of interest for
this work are shown in bold.} \label{H2COFig}
\end{figure}

As shown in Fig. \ref{H2COFig}(a), H$_{2}$CO is a near-symmetric
prolate top molecule, with nearly degenerate rotations about the b-
and c- axes. The dipole moment, $\mu_a$, along the a-axis causes
these nearly degenerate, opposite parity states to experience a
large Stark shift with an applied electric field, as shown in Fig.
\ref{H2COFig}(b-d). Here, the states are labeled by their zero-field
identity in the $|J_{\tau}m_J>$ basis, where $\tau = K_{a} - K_{c}$
($K_{i}$ is the projection of $J$ along the $i$th axis)
\cite{Hain:1999}. Of particular interest to this work is the
$|1_{1}1>$ state (Fig. \ref{H2COFig}(c)), which is the upper
component of the lowest $J$ level of ortho-formaldehyde and is hence
well populated in our supersonic expansion. This state experiences a
large Stark shift (1.32 cm$^{-1}$ at 125 kV/cm), making it excellent
for Stark deceleration. There are other states accessible via, for
example, stimulated Raman adiabatic passage that offer an improved
Stark deceleration efficiency ($|2_{2}2>$) or a good candidate for
implementing an AC Stark trap ($|0_{0}0>$) \cite{VanVeldhoven:2005}.

\textbf{OH-H$_{2}$CO Theory}. The Hamiltonian, $H$, used in the
scattering calculations takes the following form:
\begin{equation}
H = T + H_{OH} + H_{H_{2}CO} + H_{Stark} + H_{dd} + H_{sr} +
H_{chem}.
\end{equation}Here $T$ is the kinetic energy, $H_{OH}$ and $H_{H_{2}CO}$ are the Hamiltonians of the
separated molecules, and $H_{Stark}$ describes the action of the
electric field on the individual molecules. $H_{dd}$ represents the
long-range dipole-dipole interaction between the molecules, which
depends on the orientations of the molecules relative to the field
axis, as well as the distance $R$ between them. This will be
expressed in a basis of molecular eigenstates (including the effect
of the field), as was done in Ref. \cite{BOHN}. In addition, to
prevent the dipole-dipole interaction from overwhelming the
short-range interaction, we replace its $1/R^{3}$ variation with
$1/(R^{3} + c_{dd})$ for an appropriate constant $c_{dd}$
\cite{Kuhn:1999}. The chemical reaction dynamics are modeled in a
schematic way by the terms $H_{sr}$ and $H_{chem}$. The goal of
$H_{sr}$ is to mimic an anisotropic interaction with a barrier, not
necessarily to characterize the OH-H$_{2}$CO system. This
approximation is justified for the low reaction barrier where the
dipole-dipole interaction dominates over internal molecular
dynamics. Thus, we construct a potential as an expansion into
Legendre functions,\begin{equation} H_{sr} = V_{o}(R) -
2V_{2}(R)C_{20}(\theta,\phi).
\end{equation}Here $(\theta,\phi)$ represent the spherical angles of the relative coordinate
between the molecules, referred to the electric field axis, and
$C_{20}$ is a reduced spherical harmonic \cite{Brink:1993}. The
isotropic part of this potential carries the barrier at a radius $R
\approx R_{b}$ ,\begin{equation}V_{o}(R)  = \frac{C_{12}}{R^{12}} -
\frac{C_{6}}{R^{6}} + D_{B} \exp(\frac{-(R-R_{b})^{2}}{w_{b}}).
\end{equation} The height of the barrier depending on $D_B$ is fully
adjustable with $w_b$ characterizing the range. In the following, we
set the barrier height equal to the threshold energy of the incoming
molecule. The radial dependence of $V_2$ is chosen to have a
magnitude comparable to $V_o$, but to change sign as $R$ crosses
$R_{b}$. Thus the polar molecules attract in a head-to-tail
configuration for $R > R_{b}$, but attract in a tail-to-tail
configuration upon crossing the barrier. Finally, chemical
reactivity ($H_{chem}$) is modeled by coupling to a purely repulsive
artificial channel. The coupling to this channel is represented by a
decaying exponential in $R$, which is of negligible size for $R >
R_{b}$. To complete a chemical reaction, the molecules must cross
over the barrier, reverse their relative orientation, and find the
artificial channel.

\begin{figure}
\resizebox{1\columnwidth}{!}{
  \includegraphics{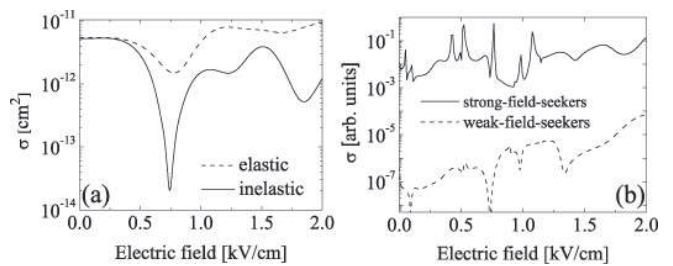}
} \caption{(a) Computed elastic and inelastic OH - H$_{2}$CO
collision cross sections as a function of external electric field at
1 mK. (b) Chemical reaction cross sections from the schematic
H-abstraction model.} \label{CrossSectionGraphs}
\end{figure}

Collision cross sections for OH-H$_{2}$CO scattering in this model
are shown in Fig. \ref{CrossSectionGraphs}(a). Here both molecules
are assumed to be in spin-stretched, weak-field-seeking states:
$|Fm_{F}, parity> = |22,f>$ for OH, and $|1_{1}1>$ for H$_{2}$CO.
The cross section for elastic (spin-changing) collisions is shown as
a solid (dashed) line. The cross sections are large and comparable,
and exhibit strong modulations as a function of an applied electric
field. This has been explained for OH-OH scattering on the basis of
purely long-range scattering by the dipole-dipole interaction
\cite{BOHN}. Thus, these results are robust and independent of the
detailed form of the short-range interaction.

The sensitive dependence of cold molecules on internal state and
electric field engenders new prospects for probing and controlling
the chemical reaction itself. To illustrate this, we show in Fig.
\ref{CrossSectionGraphs}(b) the chemical reaction cross section
versus electric field. The collision energy is assumed to be 1 mK.
(Note that, because the reaction mechanism is schematic in our model
we report only relative values for the cross sections.) The solid
line is for collisions in which both OH and H$_2$CO molecules are in
their strong-field-seeking states. At this energy the field has a
profound influence on the reaction cross section, mostly by
accessing a very large number of resonant states. Considering that
each resonant state spans a different region of configuration space,
mapping out resonances such as these would serve as an extremely
sensitive probe to transition states. The dashed trace in Fig.
\ref{CrossSectionGraphs}(b) shows the reaction rate for
weak-field-seeking states of the reactants. In addition to probing
many resonances, this cross section rises by nearly four orders of
magnitude as the field increases. This rise is also a direct
consequence of the dipole-dipole interaction. In Refs.
\cite{BOHN,Ticknor:2005} it was shown that for polar molecules in
weak-field-seeking states, the dipole-dipole interaction can
``shield'' these molecules from getting close enough together to
react chemically. Thus the cross section is strongly suppressed
relative to that for strong-field-seekers. However, as the field is
increased, the inner turning point of the relevant potential curve
moves to smaller $R$, making the shielding less effective, and
hence, the reaction more likely.

\begin{figure}
\resizebox{0.9\columnwidth}{!}{%
  \includegraphics{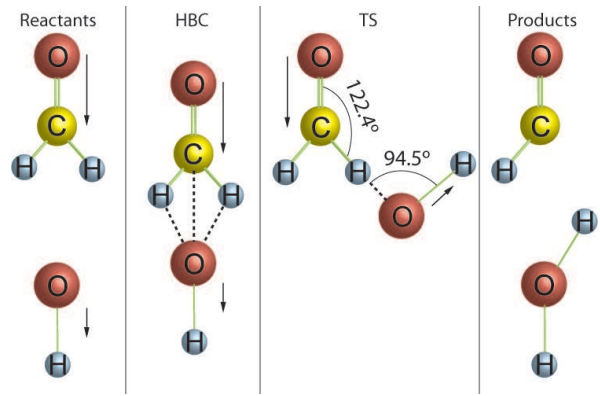}
} \caption{(Color online) The reaction mechanism for the
H-abstraction channel following Ref \cite{Dupuis:1984}. }
\label{Pathway}
\end{figure}

While the above discussion is relevant for future studies in a low
temperature trap environment, another important capability for
current experiments is the direct control of the H-abstraction
reaction barrier height through the application of an external
electric field. The basic H-abstraction reaction mechanism,
following Ref. \cite{Dupuis:1984}, is shown in Fig. \ref{Pathway}.
We note that Ref. \cite{Alvarez:2001} presents a slightly different
H-abstraction configuration, but with the same end-to-end dipole
coupling scheme. During the H-abstraction process, the energy
required to rotate one dipole moment (OH) versus the other
(H$_{2}$CO) enables our proposed control of reaction dynamics using
an external electric field. Hence, the two abstraction
configurations present no difference other than the magnitude of the
shift we can create in the reaction barrier. The important
characteristic to note is that the hydrogen-bonded complex (HBC) in
the second panel (Fig. \ref{Pathway}) forms along an attractive
direction of the two electric dipoles, but in the transition state
(TS) the OH dipole has essentially flipped its orientation relative
to the H$_{2}$CO dipole. While an external electric field, which
orients the molecules would allow (and perhaps even encourage) the
formation of the HBC; it would add an energy barrier to the
formation of the TS, and thus the H-abstraction channel through the
OH dipole-field interaction. Based on the bond geometries shown in
Ref. \cite{Dupuis:1984}, the addition to the energy barrier in going
from the HBC to the TS is $\approx 1.8\mu_{OH}|\overrightarrow{E}|$,
where $\mu_{OH}$ is the expectation value of the dipole for the OH
molecule. For a high, yet attainable, electric field of 250 kV/cm,
the additional barrier energy would be 10 K. Thus, assuming a
reaction barrier at a H$_2$CO collision speed of 223 m/s (60 K
\cite{Danna:2003}) the applied electric field would shift the
required collision velocity to 234 m/s, well within the Stark
decelerator resolution.

\textbf{Experiment}. The decelerator used for H$_{2}$CO is similar
to the apparatus in our previously described OH experiments
\cite{Bochinski:2003,Bochinski:2004,Hudson:2004}. Molecules in a
skimmed, pulsed supersonic beam are focused by an electrostatic
hexapole field to provide transverse coupling into the Stark
decelerator. The Stark decelerator is constructed of 143 slowing
stages spaced 5.461 mm apart with each stage comprised of two
cylindrical electrodes of diameter 3.175 mm separated axially by
5.175 mm and oppositely biased at high voltage ($\pm$ 12.5 kV).
Successive stages are oriented at 90$^{o}$ to each other to provide
transverse guiding of the molecular beam. The geometry of the
slowing stages provides an electric field maximum between the
electrodes with the field decreasing away from the electrode center.
Switching the electric field when the molecules are directly between
two adjacent stages (no net deceleration) is denoted by a
synchronous phase angle of $\phi_{o}= 0^{o}$ (often referred to as
bunching), while switching the electric field when the molecules are
between the electrodes (maximum deceleration) is denoted by
$\phi_{o}= 90^{o}$.

The H$_{2}$CO molecules are produced from the cracking of the
formaldehyde polymer to produce the monomer, which is passed through
a double u-tube apparatus \cite{Spence:1935}. Xenon at 2 bar
pressure is flowed over the collected H$_{2}$CO, held at 196 K where
H$_{2}$CO has $\sim$2.7 kPa (20 Torr) vapor pressure. This
Xe/H$_{2}$CO mixture is expanded through a solenoid type supersonic
valve producing a beam with a mean speed of 350 m/s with
approximately a 10\% velocity spread.

H$_{2}$CO is detected using laser-induced fluorescence. The
molecules are excited from the $|1_{1}1>$ ground state by photons at
353 nm generated from a frequency-doubled, pulsed-dye laser pumped
by Nd:YAG laser to the $\widetilde{A}^{1}$A$_{2}$ electronically
excited state with one quantum in the $\nu_4$ out-of-plane bending
vibrational mode. Approximately 40\% of the excited H$_2$CO decays
non-radiatively \cite{Henke:1982}, while the remaining molecules
emit distributed fluorescence from 353 nm to 610 nm
\cite{Shibuya:1979}.

\begin{figure}
\resizebox{1\columnwidth}{!}{%
  \includegraphics{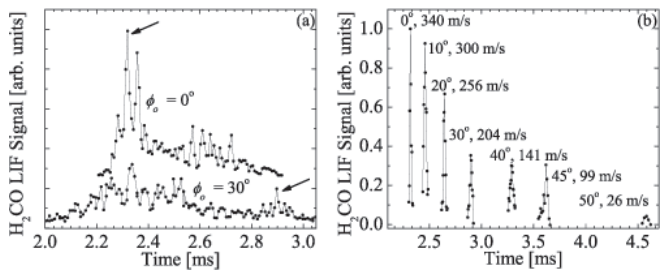}
} \caption{Stark deceleration of H$_2$CO molecules, showing (a)
bunching ($\phi_o = 0^o$) and Stark deceleration at $\phi_o = 30^o$
(offset down for clarity). (b) Stark decelerated packets of H$_2$CO
molecules at the output of the decelerator.} \label{ToFCurves}
\end{figure}

Shown in Fig. \ref{ToFCurves} are time-of-flight (ToF) measurements
taken at the exit of the decelerator, which is $\sim$0.9 m
down-stream from the valve. In these figures the time axis is
relative to the beginning of deceleration. In Fig.
\ref{ToFCurves}(a) where the decelerator is operated at $\phi_{o}=
0^{o}$, a large peak in the ToF curve denotes the arrival (marked by
an arrow) of the phase stable packet whose density is $\sim$10$^{6}$
cm$^{-3}$ and contains a few million molecules. It is important to
note that even though the speed of this packet is 340 m/s, 10 m/s
lower than the pulse mean speed, no deceleration has occurred.
Instead the molecules in the pulse whose speed was in the range from
340 m/s $\pm$ 25 m/s were loaded into the packet as detailed in our
earlier work \cite{Hudson:2004}. Also shown in Fig.
\ref{ToFCurves}(a), but offset for clarity is the ToF curve for
deceleration at $\phi_{o}= 30^{o}$.  Here the arrow denotes the
arrival of a slowed H$_2$CO packet with a mean velocity of 204 m/s.
Displayed in Fig. \ref{ToFCurves}(b) are ToF curves for the slowed
molecular packets as a function of $\phi_o$. The minimum speed shown
is 26 m/s for $\phi_o = 50^o$, resulting in a temperature of
$\sim$100 mK.

To observe the OH-H$_2$CO collision and reaction dynamics requires
monitoring both the OH population and the production of the formyl
radical (CHO). We have developed sensitive fluorescence imaging
techniques for OH \cite{Bochinski:2004}, and because CHO has a low
ionization potential (8.1 eV), high signal-to-noise ratio
photo-ionization detection of the reaction can be implemented.
Nevertheless, given the current available densities of the reactants
the observation of the dynamics proposed here will be daunting.
However, further improvements such as multiple trap loading and
secondary cooling of the molecular samples should make these
observations possible.

In summary, we have detailed new methods to study interspecies
molecular collision and reaction dynamics, and have shown how
control of these processes may be achieved. We have also
demonstrated preparation of both cold polar molecules required for
these studies. Experiments are currently poised to begin exploring
the rich physics presented by the dipolar interaction.

\begin{acknowledgments}
Funding support for this work comes from NSF, NIST, DOE, and the
Keck Foundation. CAT is supported by the Division of Chemical
Sciences, Geosciences, and Biosciences, the Office of Basic Energy
Sciences, the DOE.
\end{acknowledgments}

\bibliography{H2CO_Hudson_Bib}

\end{document}